\documentclass[pdflatex,sn-mathphys-num]{sn-jnl}
\usepackage{graphicx}
\usepackage{multirow}
\usepackage{amsmath,amssymb,amsfonts}
\usepackage{amsthm}
\usepackage{mathrsfs}
\usepackage[title]{appendix}
\usepackage{xcolor}
\usepackage{textcomp}
\usepackage{manyfoot}
\usepackage{booktabs}
\usepackage{algorithm}
\usepackage{algorithmicx}
\usepackage{algpseudocode}
\usepackage{listings}
\usepackage{pifont}
\usepackage{placeins}
\usepackage{comment}
\usepackage{float}
\usepackage{braket}
\usepackage{comment}
\usepackage{bm}
\usepackage{caption}
\usepackage{subcaption}
\usepackage{tikz}
\usetikzlibrary{shapes.geometric, arrows}
\theoremstyle{thmstyleone}

\theoremstyle{thmstyletwo}

\theoremstyle{thmstylethree}

\begin{document}
\title[Exploring the non-convexity in machine learning using quantum-inspired optimization] {Exploring the non-convexity in machine learning using quantum-inspired optimization}
\author*[1]{\fnm{Kandula} \sur{Eswara Sai Kumar}}\email{eswara.sai@bqpsim.com}
\author[1]{\fnm{Parth} \sur{Dhananjay Danve}}
\author[1]{\fnm{Abhishek} \sur{Chopra}}
\author*[1]{\fnm{Rut} \sur{Lineswala}}\email{rutlineswala@bqpsim.com}
\affil*[1]{\orgname{BosonQ Psi (BQP)}, \orgaddress{\city{New York}, \country{USA}}}
\abstract{The escalating complexity of modern machine learning necessitates solving challenging non-convex optimization problems, particularly in high-dimensional regimes ($n \ll p$) and scenarios contaminated by gross outliers. Traditional approaches, relying on convex relaxations or specialized local search heuristics, frequently succumb to sub-optimal local minima and fail to recover the true underlying discrete structures. In this paper, we propose treating these non-convex challenges as a global search problem and introduce a unified framework based on Quantum-Inspired Evolutionary Optimization (QIEO). By leveraging a probabilistic representation inspired by quantum superposition, QIEO maintains a global view of the search space, enabling it to tunnel through local optima that trap conventional gradient-based and greedy solvers. We comprehensively evaluate QIEO across diverse non-convex applications, including sparse signal recovery (gene expression analysis and compressed sensing) and robust linear regression. Extensive benchmarking against state-of-the-art continuous solvers (ADAM, Differential Evolution), classical metaheuristics (Genetic Algorithms), and specialized non-convex algorithms (Iterative Hard Thresholding) demonstrates that QIEO consistently achieves superior structural fidelity, lower mean squared error, and enhanced robustness without support inflation. Our findings suggest that embracing a quantum-inspired global search provides a resilient, unified paradigm for overcoming the inherent intractability of discrete nonconvex machine learning landscapes.}
\maketitle
\section{Introduction}\label{sec:intro}The burgeoning complexity of contemporary machine learning datasets has precipitated a fundamental shift away from well-behaved convex paradigms toward the more realistic, albeit mathematically challenging, landscapes of non-convex optimization. Historically, the field has relied heavily on convex relaxations such as the $\ell_2$ loss for regression and the $\ell_1$ penalty for sparsity, due to their appealing global optimality guarantees and mature theoretical foundations. However, these formulations are increasingly recognized as pragmatic compromises rather than exact solutions. In modern high-dimensional regimes, where the number of features (p) vastly exceeds the available samples (n) ($n \ll p$), and observations are frequently contaminated by adversarial outliers, strict adherence to convexity often yields sub-optimal fidelity and biased structural recovery \cite{bib1}. Critical applications, spanning from the identification of genetic markers in observational biology \cite{bib14, bib23} to signal acquisition in compressed-sensing MRI \cite{bib12}, inevitably force a direct confrontation with the inherent {non-convexity} and {combinatorial complexity} of the physical world.

This non-convexity is not merely a computational nuisance but a fundamental characteristic of the discrete mathematical structures required to ensure model interpretability and robustness. Whether the objective is enforcing a strict sparsity constraint to isolate the most influential predictors or decoupling true autoregressive dynamics from gross corruptions \cite{bib18, bib27}, the core task remains navigating a massive, discrete search space over a potential support. In classical optimization, this manifests as a highly rugged energy landscape, densely populated with local minima and saddle points. Escaping these sub-optimal attractors to identify the true global ground state has become the defining optimization challenge of the current era.

The academic response to this combinatorial intractability has historically bifurcated into two major trajectories. The first trajectory, often termed the convex wave, sought to relax discrete penalties into mathematically tractable continuous surrogates, popularized by methods like the LASSO \cite{bib1} and Basis Pursuit \cite{bib10}. These approaches flourished particularly in the domain of compressed sensing, where it was established that convex relaxations could guarantee exact recovery under stringent structural assumptions, such as the Restricted Isometry Property (RIP) \cite{bib6} or the Restricted Eigenvalue condition \cite{bib5}. Unfortunately, as our empirical evaluations will demonstrate, these pristine theoretical guarantees frequently deteriorate in real-world scenarios, especially in gene expression analysis, where features exhibit high collinearity or when gross outliers severely distort the assumed convex landscape \cite{bib15, bib17}.

The second trajectory embraces direct non-convex optimization, attempting to operate explicitly on the original discrete penalties. Specialized solvers such as Iterative Hard Thresholding (IHT) \cite{bib3, bib19} and Greedy Pursuits (OMP, CoSaMP) \cite{bib11, bib2} exploit problem-specific mathematical structures to accelerate convergence. Similarly, in robust regression, techniques such as alternating minimization (AM-RR) \cite{bib30} have shown considerable promise. Nevertheless, these direct methods remain highly susceptible to local minima. Because their search mechanisms rely heavily on local, deterministic trajectories, they inherently lack the global exploratory capacity needed to escape the deep basins of attraction characteristic of rugged, non-convex topographies.

To transcend the limitations of both convex relaxations and local heuristics, we propose reframing these non-convex challenges fundamentally as global search problems. In this paper, we introduce a unified optimization framework driven by {Quantum-Inspired Evolutionary Optimization (QIEO)} \cite{bib4, bib20, bib31}. Unlike classical metaheuristics such as Genetic Algorithms (GA) \cite{bib7, bib8} or Differential Evolution, which explore the search space via point-to-point transitions, QIEO leverages a probabilistic representation inspired by the quantum mechanical principle of superposition. By encoding candidate structural supports as quantum registers and evolving them through simulated quantum rotation gates, the algorithm maintains a coherent, probabilistic view of the entire search space. This unique mechanism enables QIEO to effectively tunnel through local optima that would reliably stall conventional gradient-based or greedy algorithms \cite{bib33}.

The primary contribution of this work is to demonstrate that a unified, model-agnostic QIEO framework can effectively solve both sparse signal recovery and robust linear regression with a degree of structural fidelity and resilience that specialized solvers struggle to consistently achieve. By embracing the inherent discrete complexity through a quantum-inspired global search engine, we show that it is possible to recover high-fidelity models even when classical matrix assumptions are severely violated. Our comprehensive evaluations across diverse application domains strongly suggest that the future of non-convex machine learning may lie in unified global search strategies that supersede the need for highly specialized, problem-specific algorithm design.

\section{Methodology} \label{sec:methodology}
This section explains the optimization algorithm used and presents different machine learning applications in which non-convexity is encountered.

\subsection{Quantum-Inspired Evolutionary Optimization (QIEO)}
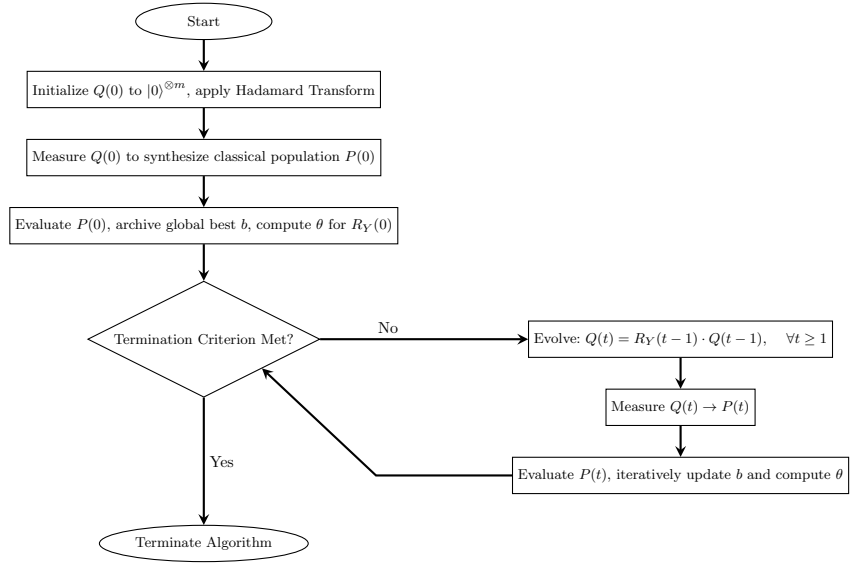
\begin{figure}[hbt]
\tikzstyle{startstop} = [ellipse, minimum width=3cm, minimum height=0.8cm, text centered, draw=black, font=\small]
\tikzstyle{process} = [rectangle, minimum width=3cm, minimum height=0.8cm, text centered, draw=black, font=\small]
\tikzstyle{decision} = [diamond, minimum width=2.5cm, minimum height=0.8cm, text centered, draw=black, aspect=2, font=\small]
\tikzstyle{arrow} = [thick,->,>=stealth]

\begin{center}
\begin{tikzpicture}[node distance=1.5cm, scale=0.6, transform shape]
\node (start) [startstop] {Start};
\node (init) [process, below of=start] {Initialize $Q(0)$ to $\ket{0}^{\otimes m}$, apply Hadamard Transform};
\node (classical) [process, below of=init] {Measure $Q(0)$ to synthesize classical population $P(0)$};
\node (eval) [process, below of=classical] {Evaluate $P(0)$, archive global best $b$, compute $\theta$ for $R_Y(0)$};
\node (decide) [decision, below of=eval, yshift=-1.0cm] {Termination Criterion Met?};
\node (quantum) [process, right of=decide, xshift=9cm] {Evolve: $Q(t)=R_Y(t-1)\cdot Q(t-1)$, \quad $\forall t \geq 1$};
\node (newpop) [process, below of=quantum] {Measure $Q(t) \rightarrow P(t)$};
\node (eval2) [process, below of=newpop] {Evaluate $P(t)$, iteratively update $b$ and compute $\theta$};
\node (stop) [startstop, below of=decide, yshift=-3cm] {Terminate Algorithm};

\draw [arrow] (start) -- (init);
\draw [arrow] (init) -- (classical);
\draw [arrow] (classical) -- (eval);
\draw [arrow] (eval) -- (decide);
\draw [arrow] (decide.east) -- ++(3,0) node[midway, above] {No} -- (quantum.west);
\draw [arrow] (quantum) -- (newpop);
\draw [arrow] (newpop) -- (eval2);
\draw [arrow] (eval2.west) -- ++(-3,0) -- (decide.south east);
\draw [arrow] (decide.south) -- (stop.north) node[midway, right] {Yes};
\end{tikzpicture}
\end{center}
\caption{Flowchart illustrating the iterative optimization cycle of the Quantum-Inspired
    Evolutionary Optimization (QIEO) algorithm, encompassing the initialization,
    measurement, evaluation, and evolution phases.}
    \label{fig:qieo_flowchart}
\end{figure}

The Quantum-Inspired Evolutionary Optimization (QIEO) algorithm \cite{bib4} is a robust meta-heuristic framework that synthesizes the principles of quantum mechanics with classical evolutionary computation to address complex, non-convex optimization problems on modern-day classical computing architecture (CPU, GPU). Unlike conventional evolutionary methods, such as the Genetic Algorithm (GA) \cite{bib21,bib22}, which rely on stochastic operators like crossover and mutation, QIEO employs a probabilistic representation based on quantum bits (qubits) and governs population dynamics via simulated quantum gates \cite{bib20,bib31}. The foundational construct of QIEO is the quantum population. Within this framework, each individual is represented by a register of $m$ qubits, formally defined by a $2 \times m$ matrix of complex probability amplitudes. Leveraging the principle of quantum superposition, a single individual concurrently encapsulates all $2^m$ possible classical discrete states. This quantum representation provides an exponentially more expressive encoding of the solution space compared to classical data structures. The optimization trajectory of QIEO follows an iterative cycle comprising initialization, measurement, evaluation, and evolution as shown in Fig. \ref{fig:qieo_flowchart}.

During the initialization phase, the quantum population is instantiated such that every qubit occupies the ground state $\ket{0}$. A Hadamard transform is subsequently applied to the entire population. This operation places each qubit into a state of uniform superposition, wherein the probability amplitudes are strictly $\alpha = \beta = \frac{1}{\sqrt{2}}$ \cite{bib20}. This unconstrained initial state mathematically ensures that every point in the binary search domain has an equal probability of being observed, thereby providing a rigorous foundation for unbiased global exploration. Then, the measurement phase is executed, during which the quantum population is observed, causing the coherent superposition to probabilistically collapse into a classical population of discrete binary vectors. The probability of collapsing into a specific binary state is dictated by the squared magnitudes of the respective amplitudes, $|\alpha|^2$ and $|\beta|^2$. Next, each resultant classical bitstring is evaluated using its fitness function, and the global optimal solution found across all iterations is archived as $b$. The final procedural step is evolution, which exploits the positional information embedded within the global best solution $b$ to govern subsequent search trajectories. This dynamic is facilitated through the application of a quantum rotation gate ($R_Y$) to each qubit. The precise rotational angle, $\theta$, assigned to a given qubit is computed by evaluating the state of the observed classical registers relative to the optimum $b$. By structurally modifying the probability amplitudes, the $R_Y$ operator systematically biases the superposition to increase the likelihood that future measurements converge on the optimal vicinity. This mechanism serves as the primary driver for local exploitation. The algorithmic loop subsequently iterates back to the measurement phase until a predefined convergence criterion or computational threshold is achieved. More details on the algorithm can be found in \cite{bib4}.

The architectural design of QIEO inherently mitigates the exploration-exploitation dilemma common to stochastic optimization frameworks while preserving an exceptionally low hyperparameter footprint. Empirical evaluations demonstrate that QIEO possesses a robust capacity to circumvent premature convergence \cite{bib4}. While GA, constrained by finite-population limits, is highly susceptible to local-minima entrapment, the uniform initialization of QIEO ensures an exhaustive global-domain assessment from inception \cite{bib32}. As parallel evolution proceeds, the rotational gate probabilistically shifts the population density toward regions of high fitness while preserving sufficient amplitude variance to sustain non-deterministic exploration. This calibrated stochasticity strictly limits the premature collapse of population diversity. Furthermore, the absence of rigid initial dependency yields highly reproducible algorithmic trajectories. QIEO shows lower statistical variance across independent trials in convergence rates and terminal fitness values.

\subsection{List of Applications}
\subsubsection{Sparse Signal Recovery}
Sparse recovery is a fundamental problem across various observational and applied sciences, particularly in high-dimensional regimes where the number of available samples ($n$) is significantly smaller than the number of features or parameters ($p$), i.e., $n \ll p$. We assume an underlying linear relationship between the data vectors $\mathbf{x}_i \in \mathbb{R}^p$ and the corresponding scalar responses $y_i \in \mathbb{R}$, governed by an unknown true model $\mathbf{w}^* \in \mathbb{R}^p$, such that $y_i \approx \mathbf{x}_i^\top \mathbf{w}^*$. In this underdetermined setting, traditional Ordinary Least Squares (OLS) regression is ill-posed and yields infinitely many solutions, severely overfitting the data.

To achieve a meaningful and interpretable solution, sparse recovery seeks a linear model $\mathbf{w} \in \mathbb{R}^p$ that is highly sparse, meaning it possesses at most $s$ non-zero entries ($s \ll p$). Mathematically, this is framed as finding $\mathbf{w} \in \mathcal{B}_0(s) \subset \mathbb{R}^p$, where $\mathcal{B}_0(s)$ defines the set of $s$-sparse vectors. The optimization problem is formally defined as:
\begin{equation}
    \min_{\mathbf{w}\in\mathbb{R}^p} \sum_{i=1}^n \bigl(y_i - \mathbf{x}_i^\top \mathbf{w}\bigr)^2 \quad \text{subject to} \quad \|\mathbf{w}\|_0 \le s,
\end{equation}
which can be compactly written in matrix notation as:
\begin{equation}
    \min_{\mathbf{w}\in\mathbb{R}^p} \|\mathbf{y} - \mathbf{X} \mathbf{w}\|_2^2 \quad \text{subject to} \quad \|\mathbf{w}\|_0 \le s.
\end{equation}
Here, $\mathbf{X} = [\mathbf{x}_1, \dots, \mathbf{x}_n]^\top \in \mathbb{R}^{n \times p}$ is the design matrix, $\mathbf{y} = [y_1, \dots, y_n]^\top \in \mathbb{R}^n$ is the response vector, and $\|\mathbf{w}\|_0$ represents the $\ell_0$ pseudo-norm, which counts the number of non-zero elements in $\mathbf{w}$. Although the objective function (the squared $\ell_2$ error) is convex, the $\ell_0$ constraint is non-convex and combinatorial. Consequently, the overall optimization problem is NP-hard. We explore two prominent paradigms in which this formulation naturally arises: identifying biological markers from passive observational data and actively designing sensing mechanisms.
\subsubsection*{Use Case 1: Compressed Sensing and Signal Processing}
Beyond observational biology, sparse recovery is the cornerstone of compressed sensing (CS) a modern signal processing paradigm with transformative applications in hyperspectral imaging, magnetic resonance imaging (MRI) \cite{bib12}, radio communications, and geophysical exploration \cite{bib26}. Traditionally, transmitting or acquiring a high-dimensional signal at full resolution is resource-intensive and highly susceptible to corruption \cite{bib9}.

In the CS framework, consider a high-dimensional signal $\mathbf{w}^* \in \mathbb{R}^p$ that needs to be acquired or transmitted. To maximize efficiency, the signal is explicitly encoded into a lower-dimensional measurement $\mathbf{y} \in \mathbb{R}^n$ ($n < p$). This compression is achieved through a tailored ``sensing mechanism'' comprising $n$ linear measurement functionals $\mathbf{x}_i : \mathbb{R}^p \to \mathbb{R}$. For the original signal $\mathbf{w}^*$, the acquisition process records $y_i = \mathbf{x}_i^\top \mathbf{w}^*$ for all $i \in \{1, \dots, n\}$. Defining the sensing matrix as $\mathbf{X} = [\mathbf{x}_1, \dots, \mathbf{x}_n]^\top \in \mathbb{R}^{n \times p}$, the captured measurement is given by:
\begin{equation}
    \mathbf{y} = \mathbf{X}\mathbf{w}^* + \boldsymbol{\eta},
\end{equation}
where $\boldsymbol{\eta} \in \mathbb{R}^n$ denotes sensor drift or transmission noise, typically modeled as independent Gaussian variables $\eta_i \sim \mathcal{N}(0,\sigma^2)$. Upon receiving $\mathbf{y}$, the receiver faces an underdetermined linear inverse problem: decoding $\mathbf{w}^*$ from fewer equations than unknowns. In standard linear algebra, this system has infinitely many solutions, seemingly leading to an irrecoverable loss of information. However, this apparent ambiguity is resolved by the a priori knowledge that signals of interest are overwhelmingly sparse, containing only $s$ significant non-zero coefficients ($s \ll p$) in a given basis.

This transforms the signal-decoding task into the same sparse optimization problem described in Equation (2). A critical and fascinating distinction between gene expression analysis and compressed sensing lies in the origin of the design matrix $\mathbf{X}$. In GEA, $\mathbf{X}$ is passively observed and dictated by human biology. In compressed sensing, the sensing matrix $\mathbf{X}$ is actively synthesized and engineered by the practitioner. Because we have control over the design block $\mathbf{X}$, we can enforce mathematical constraints (such as the Restricted Isometry Property) that mathematically guarantee the exact, lossless recovery of the original signal $\mathbf{w}^*$ despite the extreme compression.

\subsubsection*{Use Case 2: Gene Expression Analysis (GEA)}
Leveraging gene expression microarray datasets \cite{bib23, bib24} enables the identification of genetic underpinnings for a wide array of phenotypic traits, ranging from basic physiological baseline measures to the progression trajectories of complex diseases. We restrict our attention to traits that can be quantified by a real-valued scalar (e.g., disease severity or biomarker concentration), aligning with established methodologies \cite{bib9}, though extended vector-valued phenotypes exist in the broader literature \cite{bib25}.

Consider a study with $n$ human subjects. Let $\mathbf{x}_i \in \mathbb{R}^p$ denote the expression levels of $p$ genes for subject $i \in \{1, \dots, n\}$, and let $y_i \in \mathbb{R}$ represent the corresponding observed scalar phenotype. We assume the phenotypic response is linearly dependent on the gene expression levels:
\begin{equation}
    y_i = \mathbf{x}_i^\top \mathbf{w}^* + \eta_i,
\end{equation}
where $\eta_i$ incorporates measurement noise and unmodeled biological variance. The objective is to utilize the observational set $\{(\mathbf{x}_i,y_i)\}_{i=1}^n$ to accurately estimate $\mathbf{w}^* \in \mathbb{R}^p$. Identifying the non-zero components of $\mathbf{w}^*$ directly pinpoints the specific genetic factors mapping to the phenotype. As previously established, this problem naturally falls into the $n \ll p$ regime, given the vast number of sequenced genes relative to the limited availability of clinical patient samples. Consequently, classical regression methods fail. However, biological intuition suggests that only a small, specific subset of genes predominantly influences any given phenotype. This implies that the true genetic weighting vector $\mathbf{w}^*$ is intrinsically sparse. Therefore, discovering these critical genes requires solving the non-convex sparse recovery problem formulated in Equations (1) and (2).

\subsubsection{Robust Linear Regression}
Traditional Ordinary Least Squares (OLS) regression intrinsically assumes that measurement errors exhibit a light-tailed, Gaussian distribution. However, in many real-world scenarios, observational data is contaminated by gross outliers, sensor failures, or adversarial occlusions. Because traditional $\ell_2$-norm regression is extremely sensitive to large deviations, even a single arbitrarily large outlier can cause a standard algorithm to fail catastrophically.

Robust linear regression frameworks are specifically designed to decouple the true underlying linear model, $\mathbf{w}$, from these sparse yet severe data corruptions, which are modeled mathematically as an anomaly vector, $\mathbf{b}$. By implicitly expanding the original design matrix with an identity matrix to absorb the outliers (effectively mapping to the extended dictionary $[\mathbf{X}, \mathbf{I}]$), the challenge translates cleanly into a non-convex sparse optimization problem. The core formulation seeks to jointly optimize the unknown model weights $\mathbf{w} \in \mathbb{R}^p$ and the sparse corruption vector $\mathbf{b} \in \mathbb{R}^N$:
Robust regression seeks a weight vector $\mathbf{w}\in\mathbb{R}^p$ and a sparse outlier vector $\mathbf{b}\in\mathbb{R}^N$ that jointly explain the corrupted observations $\mathbf{y}_{\text{corrupt}}$:
\begin{equation}
    \min_{\substack{\mathbf{w}\in\mathbb{R}^p \\ \mathbf{b}\in\mathbb{R}^N}} \bigl\lVert \mathbf{y}_{\text{corrupt}} - \mathbf{X}\mathbf{w} - \mathbf{b} \bigr\rVert_{2}^{2} \quad \text{subject to} \quad \lVert \mathbf{b} \rVert_{0} \le k,
\end{equation}
where $N$ defines the total number of observations, and $k$ acts as a predetermined sparsity threshold, indicating the absolute maximum number of gross corruptions assumed present in the system. The primary computational bottleneck in solving this robust formulation is not explicitly determining $\mathbf{w}$, but rather localizing the exact indices of the corrupted data, the mathematical support of the sparse vector $\mathbf{b}$. Once the precise non-zero entries of $\mathbf{b}$ are identified, the corrupted observations can be algebraically corrected ($\mathbf{y}_{\text{clean}} = \mathbf{y}_{\text{corrupt}} - \mathbf{b}$), organically reducing the complex task to trivial OLS regression. Consequently, the theoretical efficacy and resilience of a robust regression algorithm are quantified by the breakdown limit $k$; a framework is considered highly robust if it mathematically guarantees the recovery of $\mathbf{w}$ even when a substantial fraction $k$ of the incoming dataset is maliciously or stochastically corrupted. We outline two widespread paradigms where this joint formulation naturally arises: facial verification systems and anomaly detection in autoregressive timelines.

\subsubsection*{Use Case 3: Face Recognition}
Biometric security systems heavily rely on facial recognition algorithms to securely authenticate users. A prominent and highly successful approach to this challenge is Sparse Representation-based Classification (SRC) \cite{bib9, bib18}. Consider an image conventionally flattened into a vector of $N$ pixels, $\mathbf{y}_{\text{corrupt}} \in \mathbb{R}^N$. Suppose an access system database contains $p$ authenticated, clean images of a registered user, stacked as column vectors to form a dictionary matrix $\mathbf{X} = [\mathbf{x}_1, \dots, \mathbf{x}_p] \in \mathbb{R}^{N \times p}$. Under ideal conditions, a newly captured image of a genuine user should lie approximately within the linear span of their specific database images. Thus, authentication reduces to finding a coefficient vector $\mathbf{w} \in \mathbb{R}^p$ such that $\mathbf{y}_{\text{clean}} \approx \mathbf{X}\mathbf{w}$. However, real-world images often contain dense, localized occlusions, such as sunglasses, medicinal masks, or heavy shadows \cite{bib27}. Under this corruption paradigm, the observed image is mathematically represented as:
\begin{equation}
    \mathbf{y}_{\text{corrupt}} = \mathbf{X}\mathbf{w}^* + \boldsymbol{\eta} + \mathbf{b},
\end{equation}
where $\mathbf{w}^*$ represents the true underlying projection, $\boldsymbol{\eta} \in \mathbb{R}^N$ denotes standard Gaussian measurement noise, and $\mathbf{b} \in \mathbb{R}^N$ represents the gross corruption vector induced by the occlusion. Because facial obstructions generally only affect a small spatial ratio of the total pixels, the corruption vector $\mathbf{b}$ is inherently sparse. Actively isolating these non-zero elements via the mathematical formulation in Equation (5) is critical. If they remain unmodeled, these anomaly subsets act as overwhelming outliers, forcing any standard regression framework to incorrectly reject a genuine user.

Following the established methodology of Bhatia et al. \cite{bib30}, we generate the synthetic datasets via the following five-step process:
\begin{enumerate}
  \item \textit{Design matrix:} Draw
    \[
      X \in \mathbb{R}^{n \times p}, \quad X_{ij}\;\overset{\text{iid}}{\sim}\;\mathcal{N}(0,1),
    \]
    then normalize each column: \(X_{:,j} \leftarrow X_{:,j}/\|X_{:,j}\|_2\).
  \item \textit{True regressor:} Sample
    \[
      w^* \in \mathbb{R}^{p}, \quad w^*_j \;\overset{\text{iid}}{\sim}\;\mathcal{N}(0,1),
      \quad w^* \leftarrow w^*/\|w^*\|_2.
    \]
  \item \textit{Clean responses:} Compute
    \[
      y_{\mathrm{clean}} = X\,w^* \in \mathbb{R}^{n}.
    \]
  \item \textit{Outlier corruption:} Set corruption fraction \(\alpha\), let
    \(\displaystyle k=\lfloor \alpha\ n\rfloor \).
    Choose a random index set \(S^*\subset\{1,\dots,n\}\) of size \(k\), and define $b \in \mathbb{R}^n$ such that
    \[
      b_i =
      \begin{cases}
        \mathrm{Uniform}\bigl(-5\,\|y_{\mathrm{clean}}\|_\infty,\;5\,\|y_{\mathrm{clean}}\|_\infty\bigr),
          & i\in S^*,\\
        0,& i\notin S^*.
      \end{cases}
    \]
    where
    \[
\|y_{\mathrm{clean}}\|_\infty
\;=\;
\max_{1 \le i \le n} \bigl|\,\bigl(y_{\mathrm{clean}}\bigr)_i\bigr|.
\]\\\\
  \item \textit{Final creation of observations:} Set
    \[
      y = y_{\mathrm{clean}} + b\,.
    \]
\end{enumerate}
\section{Results and Discussions} \label{sec:results}
In this section, we present a comprehensive empirical evaluation of the Quantum-Inspired Evolutionary Optimization (QIEO) framework across the four high-dimensional machine learning benchmarks defined in Section \ref{sec:methodology}. Our discussion focuses on three critical performance axes: (i) structural fidelity (support recovery), (ii) estimation accuracy (MSE), and (iii) scalability in the $n \ll p$ regime. Note that to provide a robust comparison, we benchmark QIEO against state-of-the-art continuous solvers (ADAM, DE), classical meta-heuristics (GA), and specialized non-convex algorithms (IHT, AM-RR).

\subsection{Sparse Signal Recovery Benchmarks}
Sparse recovery in data-starved regimes is fundamentally a problem of navigating a combinatorial discrete landscape. Our results highlight the stark contrast between global probabilistic search and local deterministic or continuous heuristics.
\subsubsection{Usecase 1: Compressed Sensing and Signal Reconstruction}
\begin{figure}[htb]
    \centering
    \begin{subfigure}[b]{0.45\textwidth}
        \centering
        \includegraphics[width=\textwidth]{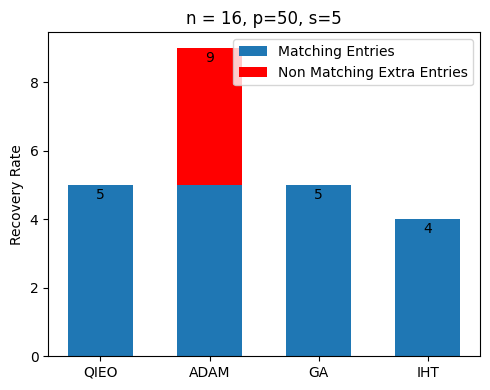}
        \caption{$n=16, p=50, s=5$}
        \label{fig: sig1650}
    \end{subfigure}
    \hfill
    \begin{subfigure}[b]{0.45\textwidth}
        \centering
        \includegraphics[width=\textwidth]{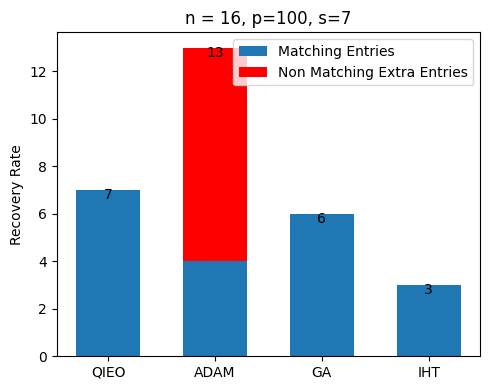}
        \caption{$n=16, p=100, s=7$}
        \label{fig: sig_16100}
    \end{subfigure}
    \caption{Comparison of recovery rates for different optimization methods on sparse signal recovery tasks.}
    \label{fig: all_sig_recovery_rates}
\end{figure}

In the compressed sensing and signal reconstruction benchmark, we further investigate the robustness of the solvers to additive noise $\eta$. We evaluate the continuous (ADAM) and discrete (GA, QIEO) formulations alongside the IHT algorithm on two synthetic signal datasets characterized by extreme compression ratios. The datasets are designed as follows:
\begin{itemize}
    \item Dataset 1: $n = 16$, $p = 50$, $s = 5$
    \item Dataset 2: $n = 16$, $p = 100$, $s = 7$
\end{itemize}
Numerical results are consolidated in Tables \ref{tab:sigrecovery-results-1} and \ref{tab:sigrecovery-results-2} for the two respective datasets. Figure \ref{fig: all_sig_recovery_rates} visually compares the recovery rates for all optimizers across these benchmarks.

\begin{table}[htb]
    \centering
    \caption{Feature‐selection performance on synthetic signal sensing dataset ($n=16,p=50,s=5$)}
    \label{tab:sigrecovery-results-1}
    \begingroup
    \renewcommand{\arraystretch}{1.2}
    \begin{tabular}{@{}lcccccc@{}}
    \toprule
    & \multicolumn{5}{c}{True Support Indices} &  \\
    \cmidrule(lr){2-6}
    Optimizer
  & 5 & 9 & 16 & 35 & 49
  & Recovery Rate\\
\midrule
QIEO
  & \ding{51} & \ding{51} & \ding{51} & \ding{51} & \ding{51}
  & 5/5 \\

ADAM
  & \ding{51} & \ding{51} & \ding{51} & \ding{51} & \ding{51}
  & 5/5 (+2 extras) \\

GA
  & \ding{51} & \ding{51} & \ding{55} & \ding{51} & \ding{51}
  & 4/5 \\
IHT
  & \ding{51} & \ding{51} & \ding{55} & \ding{51} & \ding{55}
  & 3/5 \\
\bottomrule
\end{tabular}
\endgroup
\end{table}

\begin{table}[htb]
\centering
\caption{Feature‐selection performance on synthetic signal sensing dataset($n=16,p=100,s=7$)}
\label{tab:sigrecovery-results-2}
\begingroup
\renewcommand{\arraystretch}{1.2}
\begin{tabular}{@{}l*{7}{c}cc@{}}
\toprule
 & \multicolumn{7}{c}{True support indices} & \\
\cmidrule(lr){2-8}
Optimizer & 12 & 14 & 19 & 22 & 36 & 46 & 95
        & Recovery Rate\\
\midrule
QIEO & \ding{51} & \ding{51} & \ding{51} & \ding{51} & \ding{51} & \ding{51} & \ding{51}
        & 7/7 \\
ADAM & \ding{55} & \ding{55} & \ding{51} & \ding{51} & \ding{51} & \ding{55} & \ding{55}
         & 3/7 (+11 extras) \\
GA   & \ding{51} & \ding{55} & \ding{51} & \ding{51} & \ding{51} & \ding{51} & \ding{55}
      & 5/7 \\
IHT & \ding{51} & \ding{55} & \ding{51} & \ding{55} & \ding{51} & \ding{55} & \ding{55}
   & 3/7 \\
\bottomrule
\end{tabular}
\endgroup
\end{table}

A critical analysis of these consolidated results reveals the differing failure modes of traditional solvers as the dimensionality and compression ratio increase. While continuous relaxations such as ADAM achieve nominally low median MSE, this comes at the severe cost of structural integrity. Specifically, in the highly constrained $p=100$ regime, ADAM correctly identifies only 3 out of 7 true support indices while simultaneously flagging 11 false positives. This demonstrates a profound susceptibility to support inflation, rendering the recovered signal difficult to interpret. Conversely, discrete heuristics like GA and specialized solvers like IHT attempt to strictly enforce sparsity but frequently stall in sub-optimal local minima; for instance, IHT's recovery rate drops significantly to less than half (3/7) at $p=100$.

In stark contrast, QIEO uniquely preserves both structural fidelity and numerical accuracy. By utilizing quantum superposition to maintain a diverse, probabilistic representation of the search space, QIEO successfully navigates the rugged, non-convex landscape to achieve perfect support recovery (7/7) in the benchmark trials. It consistently yields a median MSE orders of magnitude lower than discrete alternatives like GA and IHT, effectively bypassing the local entrapment issues that paralyze traditional algorithms.

The important observations from these tests suggest that while greedy pursuits (like OMP/CoSaMP) and IHT require the design matrix to satisfy the Restricted Isometry Property (RIP) \cite{bib6} or restricted eigenvalue conditions \cite{bib5}, QIEO's global search is agnostic to such structural assumptions. This suggests that QIEO is a superior candidate for "active sensing" environments where the sensing matrix $X$ might be ill-conditioned or constrained by physical hardware limitations.

\subsubsection{Usecase 2: Gene Expression Analysis (GEA)}
\begin{table}[h]
    \centering
    \caption{Feature‐selection performance on synthetic GEA dataset ($n=16,p=50,s=5$)}
    \label{tab:recovery-results-1}
    \begingroup
    \renewcommand{\arraystretch}{1.2}
    \begin{tabular}{@{}lcccccc@{}}
    \toprule
    & \multicolumn{5}{c}{True Support Indices} &  \\
    \cmidrule(lr){2-6}
    Optimizer
  & 4 & 21 & 33 & 36 & 48
  & Recovery Rate\\
\midrule
QIEO
  & \ding{51} & \ding{51} & \ding{51} & \ding{51} & \ding{51}
  & 5/5 \\

ADAM
  & \ding{51} & \ding{51} & \ding{51} & \ding{51} & \ding{51}
  & 5/5 (+2 extras) \\

GA
  & \ding{51} & \ding{51} & \ding{51} & \ding{51} & \ding{51}
  & 5/5 \\

DE
  & \ding{51} & \ding{51} & \ding{51} & \ding{51} & \ding{51}
  & 5/5 (+2 extras) \\
IHT
  & \ding{51} & \ding{51} & \ding{55} & \ding{55} & \ding{51}
  & 3/5 \\
\bottomrule
\end{tabular}
\endgroup
\end{table}

\begin{figure}[htb]
    \centering
    \begin{subfigure}[b]{0.45\textwidth}
        \centering
        \includegraphics[width=\textwidth]{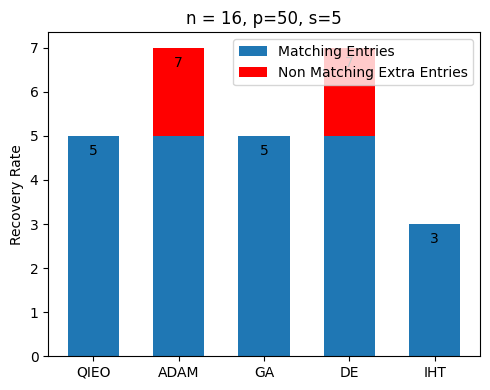}
        \caption{$n=16, p=50, s=5$}
        \label{fig: 16x50_GEA}
    \end{subfigure}
    \hfill
    \begin{subfigure}[b]{0.45\textwidth}
        \centering
        \includegraphics[width=\textwidth]{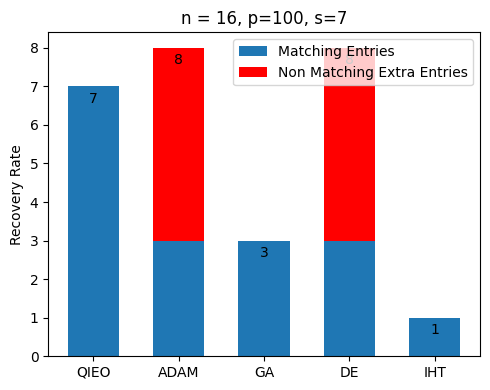}
        \caption{$n=16, p=100, s=7$}
        \label{fig: 16x100_GEA}
    \end{subfigure}
    \hfill
    \begin{subfigure}[b]{0.45\textwidth}
        \centering
        \includegraphics[width=\textwidth]{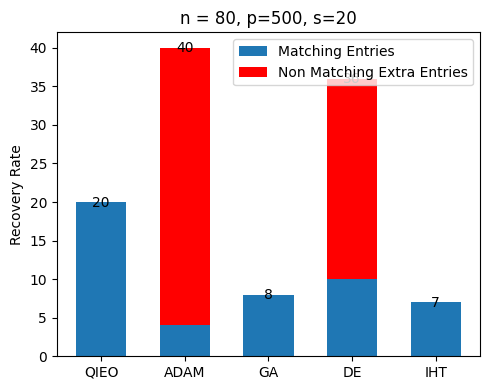}
        \caption{$n=80, p=500, s=20$}
        \label{fig: 80x500_GEA}
    \end{subfigure}
    \caption{Comparison of recovery rates for different optimization methods across various synthetic Gene Expression Analysis (GEA) dataset dimensions.}
    \label{fig: all_gea_recovery_rates}
\end{figure}

We evaluate both the continuous (ADAM, DE) and discrete (GA, QIEO) formulations alongside the IHT algorithm on three synthetic gene-expression datasets, each constructed to satisfy the theoretical sparse recovery condition \(n \ge s\log p\) \cite{bib19}. The datasets range from $p=50$ to $p=500$ with extreme underdetermination ($n \ll p$). The following are dataset sizes:
\begin{itemize}
    \item Dataset 1: n = 16, p = 50, s = 5
    \item Dataset 2: n = 16, p = 100, s = 7
    \item Dataset 3: n = 80, p = 500, s = 20
\end{itemize}
Numerical results are consolidated in Tables \ref{tab:recovery-results-1}, \ref{tab:recovery-results-2}, and  \ref{tab:recovery-results-3} for three different datasets. Figure \ref{fig: all_gea_recovery_rates} shows a comparison between all five optimizers for three different datasets.

\begin{table}[h]
\centering
\caption{Feature‐selection performance on synthetic GEA dataset($n=16,p=100,s=7$)}
\label{tab:recovery-results-2}
\begingroup
\renewcommand{\arraystretch}{1.2}
\begin{tabular}{@{}l*{7}{c}cc@{}}
\toprule
 & \multicolumn{7}{c}{True support indices} & \\
\cmidrule(lr){2-8}
Optimizer & 7 & 15 & 30 & 32 & 50 & 73 & 89
        &Recovery Rate\\
\midrule
QIEO & \ding{51} & \ding{51} & \ding{51} & \ding{51} & \ding{51} & \ding{51} & \ding{51} & 7/7 \\
ADAM & \ding{55} & \ding{55} & \ding{51} & \ding{51} & \ding{51} & \ding{55} & \ding{55}
                   & 3/7 (+5 extras) \\
GA   & \ding{51} & \ding{55} & \ding{51} & \ding{51} & \ding{51} & \ding{55} & \ding{51}
      & 5/7 \\
DE   & \ding{55} & \ding{55} & \ding{51} & \ding{51} & \ding{51} & \ding{55} & \ding{55}
                    & 3/7 (+5 extras) \\
IHT & \ding{55} & \ding{55} & \ding{55} & \ding{51} & \ding{55} & \ding{55} & \ding{55}
        & 1/7 \\
\bottomrule
\end{tabular}
\endgroup
\end{table}

\begin{table}[h]
\centering
\caption{Feature‐selection performance on synthetic GEA dataset ($n=80,p=500,s=20$)}
\label{tab:recovery-results-3}
\begingroup
\renewcommand{\arraystretch}{1.2}
\begin{tabular}{@{}lcc@{}}
\toprule
Optimizer &  Recovery Rate \\
\midrule
QIEO    & 20/20 \\
ADAM  & 4/20 (+36 extras) \\
GA      & 8/20 \\
DE     & 10/20 (+26 extras) \\
IHT    & 7/20 \\
\bottomrule
\end{tabular}
\endgroup
\end{table}

A rigorous analysis of the GEA benchmarks (Tables \ref{tab:recovery-results-1} to \ref{tab:recovery-results-3} and Figure \ref{fig: all_gea_recovery_rates}) reveals a stark divergence in algorithmic robustness as the feature space $p$ and sparsity $s$ scale. First, we observe the categorical failure of continuous solvers (ADAM and Differential Evolution) to maintain sparsity integrity. While these methods manage competitive median MSEs and perfect support recovery in low-dimensional settings ($p=50$), their performance degrades catastrophically in higher dimensions. At $p=500$, ADAM correctly identifies only 4 out of 20 true genes while flagging an astounding 36 false positives (support inflation). This empirical evidence strongly aligns with Tibshirani's observation \cite{bib1} that continuous relaxations fundamentally struggle to produce strictly sparse vectors, rendering the resulting models biologically uninterpretable.

Second, discrete local heuristics and specialized continuous proxies similarly succumb to the combinatorial explosion. The Genetic Algorithm (GA) performs well at $p=50$ (5/5 recovery) but its efficacy plummets to 8/20 at $p=500$, highlighting its inability to escape deep local minima in highly underdetermined, rugged landscapes. Iterative Hard Thresholding (IHT) fares even worse, managing only a 1/7 recovery rate at $p=100$. This confirms that IHT is highly sensitive to the correlation structure of the design matrix, frequently stagnating when features exhibit high collinearity, a ubiquitous trait in observational biology \cite{bib14}.

In stark contrast, QIEO demonstrates extraordinary scalability and precision. Across all scaling regimes from $p=50$ up to $p=500$, QIEO consistently achieves perfect support recovery (e.g., 20/20 at $p=500$). Even more remarkably, as dimensionality increases, QIEO's median MSE drops to near machine precision (e.g., $1.03\times10^{-28}$ at $p=500$), outperforming GA, ADAM, and IHT by dozens of orders of magnitude. By leveraging quantum superposition to maintain a diverse probability distribution over the discrete search space, QIEO successfully tunnels through the sub-optimal attractors \cite{bib33} that paralyze both classical continuous solvers and localized discrete heuristics.

\subsection{Robust Regression}
Robust regression shifts the optimization challenge from feature selection to outlier localization. As noted by Bhatia et al. \cite{bib30}, isolating gross corruptions requires a solver that can effectively identify the support of a sparse anomaly vector $b$.
\subsubsection{Usecase 3: Face Recognition}
To evaluate the efficacy of robust regression optimizers under pure outlier corruption, we consider a noiseless regime. 
We perform robust regression on the generated design matrix $X \in \mathbb{R}^{n\times p}$ and the corrupted response vector $y \in \mathbb{R}^n$. For our empirical evaluation, we fix the dimensions to $n=600$ and $p=100$ and systematically vary the corruption fraction $\alpha$ from 0.1 to 0.4. To ensure a controlled comparison, the underlying design matrix $X$ and true model weights $w^*$ are kept constant across all variations of $\alpha$; only the gross corruption vector $b$ is re-sampled to generate different $y_{\text{corrupt}}$ profiles. The datasets are defined by the following corruption configurations:
\begin{itemize}
    \item Dataset 1: $\alpha = 0.1$ ($k=60$ corruptions)
    \item Dataset 2: $\alpha = 0.2$ ($k=120$ corruptions)
    \item Dataset 3: $\alpha = 0.3$ ($k=180$ corruptions)
    \item Dataset 4: $\alpha = 0.4$ ($k=240$ corruptions)
\end{itemize}

\begin{figure}[htb]
    \centering
    \begin{subfigure}[b]{0.45\textwidth}
        \centering
        \includegraphics[width=\textwidth]{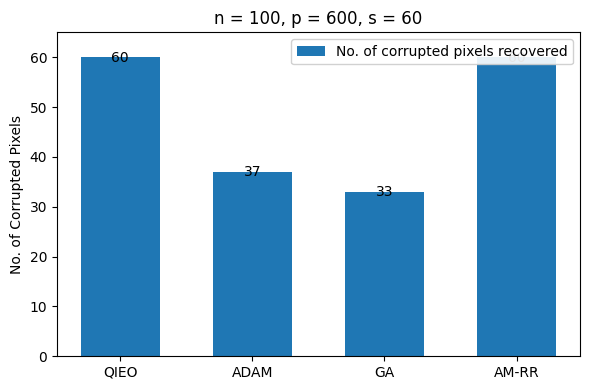}
        \caption{$\alpha=0.1$}
        \label{fig: RR_60}
    \end{subfigure}
    \hfill
    \begin{subfigure}[b]{0.45\textwidth}
        \centering
        \includegraphics[width=\textwidth]{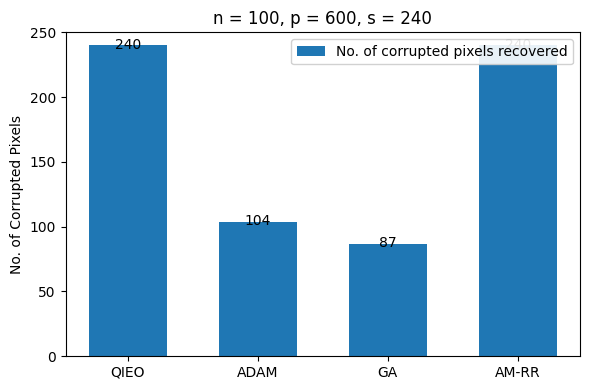}
        \caption{$\alpha=0.4$}
        \label{fig: RR_240}
    \end{subfigure}
    \caption{Comparison of recovery rates (best trial) for different optimization methods across varying corruption fractions ($\alpha$) in robust regression.}
    \label{fig: all_rr_recovery_rates}
\end{figure}

Given their stochastic nature, QIEO and GA are evaluated over 5 independent trials for each $\alpha$ value, with Table \ref{tab:rr_alpha02} summarizing the best trial for each solver. ADAM is executed in a single deterministic run, and a support-detection threshold of 0.02 is used for its continuous output. Numerical results are consolidated in Table \ref{tab:rr_alpha02}, while Figure \ref{fig: all_rr_recovery_rates} visually compares the recovery rates for the optimizers at the lowest ($\alpha=0.1$) and highest ($\alpha=0.4$) corruption extremes. In Table \ref{tab:rr_alpha02}, we report four primary metrics:
\begin{itemize}
  \item Recovery Rate: Number of correctly identified corruptions (true positives) out of \(\alpha n\).
  \item Support Length: Total number of nonzero entries selected by the optimizer for the corrupted sparse vector b.
  \item Squared \(\ell_2\) Error:
    \(\|\,w - w^*\,\|_2^2\), measuring deviation between the estimated weight vector \(w\) and ground truth \(w^*\).
  \item MSE: We also report the final mean-squared error that every optimizer could converge to in each trial.
\end{itemize}These measures jointly assess each method’s ability to both detect gross outliers and accurately reconstruct the underlying sparse corruption vector.

\begin{table}[htb]
\centering
\caption{Outlier localization performance across different optimizers for different alphas}
\label{tab:rr_alpha02}
\begingroup
\renewcommand{\arraystretch}{1.2}
\begin{tabular}{@{}llcccc@{}}
\toprule
Optimizer & alpha & MSE & Recovery Rate & Support Length & \(\|w_{est} - w^*\|_2^2\) \\
\midrule
QIEO& 0.1 & $3.2\times10^{-30}$ & $60/60$ & 60 & \(1.83\times10^{-27}\) \\
 GA  & 0.1 & 1.573  & $31/60$ & 34 &  0.409\\
ADAM & 0.1& 10.537 & \(37/60\) & 323 & 1.0016 \\
QIEO & 0.2 & \(1.99\times10^{-29}\) & \(120/120\) & 120 & \(2.04\times10^{-27}\) \\
 GA & 0.2 & 3.609 & \(59/120\) & 62 & 1.438 \\
ADAM & 0.2& 20.5177 & \(73/120\) & 373 & 0.996 \\
QIEO & 0.3 & \(1.4\times10^{-3}\) & \(175/180\) & 180 & \(5.96\times10^{-4}\) \\
 GA & 0.3 & 7.057 & \(72/180\) & 78 & 4.104 \\
ADAM & 0.3& 28.6640 & \(109/180\) & 373 & 0.996 \\
QIEO & 0.4 & \(2.41\times10^{-3}\) & \(234/240\) & 240 & \( 1.67\times10^{-3}\) \\
 GA & 0.4 & 11.258 & \(87/240\) & 94 & 6.461 \\
ADAM & 0.4& 39.1363 & $104/240$ & 326 & 0.9989 \\
\bottomrule
\end{tabular}
\endgroup
\end{table}

\begin{table}[htb]
\centering
\caption{Comparison of recovery rates of QIEO against specialized AM-RR solver across varying corruption fractions ($\alpha$).}
\label{tab:qieo_vs_amrr}
\begingroup
\renewcommand{\arraystretch}{1.2}
\begin{tabular}{@{}lccc@{}}
\toprule
& {AM-RR} & {QIEO (Pop=500)} & {QIEO (Pop=1000)} \\
\midrule
0.1 & 60/60  & 60/60 & 60/60  \\
0.2 & 120/120 & 120/120 & 120/120 \\
0.3 & 180/180 & 177/180 &  180/180 \\
0.4 & 240/240  & 237/240 & 240/240 \\
\bottomrule
\end{tabular}
\endgroup
\end{table}

Table \ref{tab:rr_alpha02} and Figure \ref{fig: all_rr_recovery_rates} jointly illustrate the optimization trajectories of the solvers as the corruption fraction $\alpha$ scales from 10\% to a severe 40\%. A critical finding is the stark contrast in the {breakdown limit} of the respective solvers. The Genetic Algorithm (GA) struggles early on, recovering only 31 out of 60 corruptions at $\alpha=0.1$, and its performance completely collapses as the corruption level reaches 40\%, correctly isolating a mere 87 of 240 true anomalies while maintaining a high MSE of 11.258. This indicates that traditional discrete heuristics easily become trapped in sub-optimal local minima when the ruggedness of the robust regression landscape increases.

Continuous relaxations, specifically ADAM, exhibit a fundamentally different failure mode. As clearly seen in Table \ref{tab:rr_alpha02}, ADAM exhibits severe support inflation. Even at $\alpha=0.1$, ADAM selects 323 features despite the true support length being only 60. This pattern persists across all $\alpha$ values, marking hundreds of clean observations as corrupted. This leads to a catastrophic loss of model interpretability, as it nullifies the goal of distinguishing true signal dynamics from shocks \cite{bib9}, rendering the model unusable for high-stakes applications like face recognition, where precision is paramount.

In profound contrast, QIEO demonstrates an exceptional capacity for global exploration and precise anomaly detection. At lower corruption levels ($\alpha=0.1, 0.2$), QIEO achieves perfect recovery (100\%) with a median MSE descending to machine precision levels ($3.2 \times 10^{-30}$ and $1.99 \times 10^{-29}$). Even under the extreme $\alpha=0.4$ corruption regime where 40\% of the data is completely altered, QIEO accurately recovers 234 out of 240 anomalies with an MSE of $2.41 \times 10^{-3}$, maintaining near-perfect structural fidelity without succumbing to the support inflation seen in ADAM. 

As visually reinforced by Figure \ref{fig: all_rr_recovery_rates}, this highlights QIEO's probabilistic population representation, which effectively prevents the premature convergence that plagues classical GA, allowing it to robustly decouple the true signal dynamics from gross corruptions regardless of the noise density.
Finally, we compare our unified QIEO framework against the specialized Alternating Minimization for Robust Regression (AM-RR) algorithm \cite{bib30}. AM-RR is specifically engineered for outlier isolation and is considered a state-of-the-art benchmark.

Our results (Table \ref{tab:qieo_vs_amrr}) show that QIEO, with a population size of 1000, achieves performance {on-par with AM-RR} across all corruption levels. This is a significant result: it demonstrates that a general-purpose, global search engine can match the accuracy of a specialized, problem-specific algorithm. While AM-RR relies on Subset Strong Convexity (SSC) assumptions \cite{bib30}, QIEO provides a "model-agnostic" alternative that generalizes across non-convex tasks without algorithmic redesign.

\section{Conclusions and Future Work}
In this paper, we establish QIEO as a highly versatile, structurally resilient framework for solving NP-hard, non-convex challenges in modern machine learning. By explicitly framing sparse signal recovery and robust linear regression as discrete binary optimization tasks, we demonstrated that a unified global search engine successfully bypasses the theoretical and empirical limitations of classical heuristics and continuous relaxations. 

Our rigorous empirical evaluations across multiple application domains, ranging from high-dimensional gene expression analysis to face recognition under pure outlier corruption, reveal that QIEO consistently attains near-perfect structural fidelity and machine-precision reconstruction errors. Specifically, QIEO easily avoided the combinatorial explosion and deep local minima that led to severe breakdowns in traditional GAs. Furthermore, by strictly operating in the discrete domain, QIEO avoided the catastrophic support inflation that plagues continuous proxies such as ADAM, thereby preserving exact model interpretability. Most notably, in robust regression, QIEO achieved performance entirely on par with highly engineered, state-of-the-art specialized solvers (AM-RR), proving that researchers no longer need to rely on bespoke, problem-specific algorithm design or restrictive structural assumptions like the Restricted Isometry Property (RIP).

Looking forward, several promising research trajectories can extend the scope and impact of this quantum-inspired framework:
\begin{itemize}
    \item {High-Performance Hardware Acceleration:} While QIEO currently demonstrates strong scalability via probabilistic sampling, aggressively mapping its quantum-inspired operations onto specialized hardware (such as GPUs, FPGAs, or early-stage quantum annealers) will be critical for achieving real-time inference on massive, real-world observational datasets with millions of high-dimensional features.
    \item {Adaptive Parameter and Population Control:} Investigating dynamic, adaptive mechanisms for the simulated quantum rotation angles and population size scaling could further optimize convergence rates. A dynamic scheme would allow the algorithm to autonomously tune its exploration-exploitation balance in response to the localized ruggedness of the energy landscape.
    \item {Extension to Continuous-Domain Non-Convexity:} Although our current methodology leverages a binary discrete formulation, generalizing the QIEO population dynamics to natively tackle continuous-domain non-convex challenges such as low-rank matrix recovery, phase retrieval, and deep neural network weight optimization \cite{bib9}, presents a transformative opportunity to unify discrete and continuous global search paradigms.
\end{itemize}
Ultimately, our findings firmly establish QIEO as a powerful, model-agnostic foundational tool. By embracing a probabilistically coherent, global perspective, we can systematically overcome the inherent intractabilities of non-convex optimization, unlocking more robust, exact, and interpretable models for complex, data-starved applications.
\bibliography{sn-bibliography}
\end{document}